\documentclass[fleqn,usenatbib]{mnras}

\usepackage{newtxtext,newtxmath}

\usepackage[T1]{fontenc}

\DeclareRobustCommand{\VAN}[3]{#2}
\let\VANthebibliography\thebibliography
\def\thebibliography{\DeclareRobustCommand{\VAN}[3]{##3}\VANthebibliography}

\usepackage{graphicx}	
\usepackage{amsmath}	
\usepackage{amssymb}	

\title[Lensing effect on BHMF]{Impact of gravitational lensing on black hole mass function inference with third-generation gravitational wave detectors}

\author[He et al.]{
Xianlong He,$^{1}$
Kai Liao,$^{1}$\thanks{liaokai@whu.edu.cn}
Xuheng Ding$^{2}$,
Lilan Yang$^{2}$,
Xudong Wen$^{3}$,
Zhiqiang You$^{3}$,
and Zong-Hong Zhu$^{1,3}$\thanks{zhuzh@whu.edu.cn}
\\
$^{1}$School of Physics and Technology, Wuhan University, Wuhan 430072, China\\
$^{2}$Kavli Institute for the Physics and Mathematics of the Universe (Kavli IPMU, WPI), University of Tokyo, Chiba 277-8583, Japan\\
$^{3}$Department of Astronomy, Beijing Normal University, Beijing, 100875, China\\
}

\date{Accepted XXX. Received YYY; in original form ZZZ}

\pubyear{2022}

\begin{document}
\label{firstpage}
\pagerange{\pageref{firstpage}--\pageref{lastpage}}
\maketitle

\begin{abstract}

The recent rapid growth of the black hole (BH) catalog from gravitational waves (GWs), has allowed us to study the substructure of black hole mass function (BHMF) beyond the simplest Power-Law distribution.
However, the BH masses inferred from binary BH merger events, may be systematically 'brightened' or 'dimmed' by the gravitational lensing effect.
In this work, we investigate the impact of gravitational lensing on the BHMF inference considering the detection of the third-generation GW detector -- the Einstein Telescope (ET).
We focus on high redshift, $z=10$, in order to obtain the upper limits of this effect.
We use Monte Carlo (MC) method to simulate the data adopting 3 original BHMFs under Un-Lensed and Lensed scenarios, then recover the parameters of BHMFs from the mock data, and compare the difference of results, respectively.
We found that all the parameters are well recovered within one standard deviation(std., 1$\sigma$), and all 3 BHMF models are reconstructed within 68\% credible interval, suggesting that lensing would not change the main structure drastically, even at very high redshifts and with high precision of ET.
And the modest influence beyond $50M_{\odot}$, depends on the modeling of the high mass tail or substructure of BHMF.
We conclude that the impact of lensing on BHMF inference with ET can be safely ignored in the foreseeable future.
Careful handling of lensing effects is required only when focusing on an accurate estimation of the high mass end of BHMF at high redshifts.

\end{abstract}


\newpage 

\section{Introduction}
\label{sec:introduction}

Since the LIGO-Virgo Collaboration announced the first binary black hole (BBH) merger in February 2016 (\citealt{LSC2016PRL}, and its basic physics explanation \citealt{LSC2016AnP} as a pedagogical introduction), 90 gravitational wave (GW) events have been accumulated in this past five years \citep{LSC202111GWTC3}. They open up a new window for astrophysics.
Studying population property is one of the applications \footnote{\label{foot:LSCpaper}\url{https://www.ligo.caltech.edu/page/detection-companion-papers}}.
The Black Hole Mass Function (BHMF), especially its details or substructures, is one of the important parts of the population property\citep{LSC202111_GWTC3_pop}.
 
The BHMF is vital to distinguish the theoretical models of stellar evolutionary channels \citep[see Intro. of][for a summary]{LSC202105_GWTC2_pop}. 
The growing catalogue of BHs from GWs helps us model and estimate it fairly accurately. 
However, the current gravitational wave detector network LIGO-VERGO-KAGRA still has considerable uncertainties \citep{Ghosh2016PhRvD..94j4070G} about the masses of the BHs estimated from the BBH-GW events, preventing us from precisely estimating the BHMF.
This instrumental noise would inevitably make some apparent high mass outlier event, 
such as {\tt GW170729}\citep{Fishbach2020}.
The next generation GW detectors such as ET, CE, DECIGO, are expected to observe more black hole events with higher accuracy and longer distances \citep{Vitale2017PhRvD..95f4052V}, to promote a better understanding of the BHMF properties.

Besides the instrument noise, gravitational lensing is another important source of the systematic uncertainty of estimation of BBH mass.
Intrinsically faint GW events at high redshift can be magnified by gravitational lensing \citep[][]{Wang1996PhRvL..77.2875W}, just as the traditional electromagnetic(EM) wave events \citep{Treu2010ARA&A..48...87T, Oguri2019Rep.Prog.Phys.82,2022arXiv220713489L,Schneider1992LensesBook,ABC2018Lens}.
Therefore, the lensing effect can bias BH mass estimation inferred from GW events, serving as an extra component of the uncertainty on the BHMF.
 
Recently, many works have investigated the influence of gravitational lensing on gravitational waves, especially for strongly lensed GWs.
In the case of the 2nd generation GW detector, the number of strongly lensed GW event is low, i.e, only a few cases per year could be detected \citep{2018PhRvD..97b3012N,2021ApJ...921..154W}.
This number is predicted to increase to hundreds per year (\citealt{Li2018MNRAS.476.2220L}, \citealt{Biesiada2014JCAP...10..080B,Ding2015JCAP...12..006D,yang2022MNRAS.509.3772Y}), when the 3rd-generation gravitational wave detectors land and run as expected in the next ten years, such as Einstein Telescope (ET) (\citealt{Sathyaprakash2012CQGra..29l4013S,Maggiore2020JCAP...03..050M}, and other documents\footnote{\label{foot:ET}\url{http://www.et-gw.eu/index.php/relevant-et-documents} important docs, and list \url{https://apps.et-gw.eu/tds}}).
Most of the lensing events are contributed from BBH events,
which could be used for testing gravity\citep{2017PhRvL.118i1101C,Fan2017PhRvL.118i1102F,2021PhRvD.103b4038G}, probing dark matter\citep{2022A&A...659L...5C} and precision cosmology\citep{liao2017NatCo...8.1148L}.
It was found these strongly- and weakly-lensed GW events could deviate the distribution of the BBH mergers at the high mass end in the observer frame, i.e.,  brightening helps to detect dimmer events under the threshold and vice versa \citep{Oguri2018MNRAS.480.3842O, Dai2017PhRvD..95d4011D}.
\citet[]{Shan2021MNRAS.508.1253S} used a semi-analytical approach to calculate the average magnification and uncertainty as a function of observed luminosity distance: ${\langle \mu \rangle, \sigma} \sim d_{obs}$.

The lensing effect was ignored in GWTC-2 \citep{LSC202105_GWTC2_pop},
because the lensing rate is rather low at the redshift we could observe right now.
The LIGO-Virgo Collaboration performed a comprehensive lensing analysis of both GWTC-1 and GWTC-2
\citep{Hannuksela2019ApJ...874L...2H, LSC2021LensAPJ},
following the strategy proposed by \citet{Haris2018arXiv180707062H}, 
but found no sufficient evidence for lensing pairs.

However, the lensing effect maybe non-negligible for the next-generation GW detectors, owing to the improved sensitivity, e.g., ET is predicted to capture farther event up to redshift $z\sim30$ \citep[][]{Maggiore2020JCAP...03..050M}.
In this work, we investigate the impact of the (strong+weak) gravitational lensing on the intrinsic BHMFs, in the context of ET.
We focus on the intrinsic BHMF at the source frame with observational uncertainties, 
while theoretical event rate distribution at the detector frame was discussed in \citet[]{Oguri2018MNRAS.480.3842O, Dai2017PhRvD..95d4011D}.
We adopt the methodology of evaluating the estimation accuracy of BHMFs from \citet{Ding_2020}.
 
This paper is organized as follows. The data simulations are introduced in Section \ref{sec:data}, for both Un-Lensed and Lensed scenarios.
We give BHMF inferences for both cases in Section \ref{sec:estimation}.
The results are compared in Section \ref{sec:result}.
We summarize in Section \ref{sec:summary}. 
And in Appendix.\ref{sec:convolution}, we present an additional perspective as a comparison.
The units of the parameters characterizing the mass (i.e. $m_\text{up}, m_\text{low}, \delta_m, \mu_p, \sigma_p$) are all $M_{\odot}$, where $M_{\odot}$ is the Solar Mass, and no units for the other parameters.
Throughout the work, a standard $\Lambda$CDM model \citep[][]{Planck2018Cosmo_Parameters} is assumed with $H_0=70\text{km}\cdot\text{s}^{-1}\cdot\text{Mpc}^{-1},\Omega_{\text{m}} = 0.30,\Omega{_\Lambda} = 0.70$.

\section{Data simulation}
\label{sec:data}
In this section, we introduce how to generate the mock catalogue of lensed GW under ET detection scenario.
In Sec. \ref{sec:catalog generation}, we generate a set of simulated BBH merger catalogs (400,000 events in our work) in the source frame to mock real universe,  using Monte Carlo (MC) sampling method.
We adopt 3 different BHMFs, i.e., truncated power law, power law with peak, and broken power law, selected from \citet{LSC202105_GWTC2_pop}, see details in Sec.\ref{sec:mass functions}
Then in Sec. \ref{sec:lensing}, we apply lensing effect on these mock events.
Note that, We assume the influence of lensing could be totally represented by the magnification of the source masses estimated by the GW signals. 
Finally, we consider the measurement uncertainties in a simple form of Log-Normal distribution following \citet[]{Ding_2020} in Sec. \ref{sec:noise}.

\subsection{Catalog Generation}
\label{sec:catalog generation}

A sufficient number of BBHs in catalogue is built by generating the initial parameters of each BBH event through MC methods.
But there are up to 15 parameters to be estimated in GW pipeline {\tt Baystar}  or {\tt PyCBC} code \citep{SingerPrice2016PhysRevD.93.024013}.
So for simplicity, we only consider several key parameters following \citet{Ding_2020}, such as: the source redshift $z$ or the luminosity distance $D_L(z)$, four angles $(\theta,\phi,\psi,\iota)$ or their combination $\Theta$, describing the BBH system's direction of the center of mass and angular momentum relative to our detector, and the respective masses of the two black holes $(m_1,m_2)$ with their chirp mass $\mathcal{M}$ before their merger.

These key parameters together form the signal-to-noise ratio(SNR) $\rho$, which represents the flux (or energy) carried by the GW signal:
\begin{equation}
\label{eq:SNRrho}
    \rho = 8\Theta \frac{r_0}{D_L(z)}\left( \frac{(1+z)\mathcal{M}}{1.2M_{\odot}} \right)^{\frac{5}{6}}\sqrt{\zeta(f_\text{max})}.
\end{equation}
A detection threshold $\rho_\text{th}$ is assumed \citep[][]{Finn1993PhysRevD.47.2198,Taylor&Gair2012PhysRevD.86.023502}, below which the GW signals are too weak to be detected.
Here $r_0=1527\text{ Mpc}$ \citep[see][Table I]{Taylor&Gair2012PhysRevD.86.023502} is the characteristic distance parameter of the detector for ET and $\zeta(f_\text{max})$ is the dimensionless function reflecting the overlap between the GW signal and the ET’s effective bandwidth. 
For simplicity, we follow the approximation in \citet{Taylor&Gair2012PhysRevD.86.023502} :$\zeta(f_\text{max})=1$ . 
The chirp mass $\mathcal{M}$ is defined as $\mathcal{M}:=\frac{(m_1m_2)^{3/5}}{(m_1+m_2)^{1/5}}$. 
The luminosity distance $D_L(z)$ is calculated in a certain cosmology model as Eq.\ref{eq:Dl-z}.

The orientation factor $\Theta$ is composed of  four orientation angles $(\theta,\phi,\psi,\iota)$. 
The first two denote the direction of the observatory on the earth relative to the line-of-sight from BBH, just the same as in the 2D spherical coordinate. 
The last two represent the direction of the orbital angular momentum of the BBH relative to the line-of-sight. 
They are independent and spherical. 
So we could assume that all of them satisfy the uniform distribution $(\cos{\theta},\phi/\pi,\psi/\pi,\cos{\iota})\in[-1,1]$, and do MC sampling.

\begin{figure}
    \centering
    \includegraphics[width=\columnwidth]{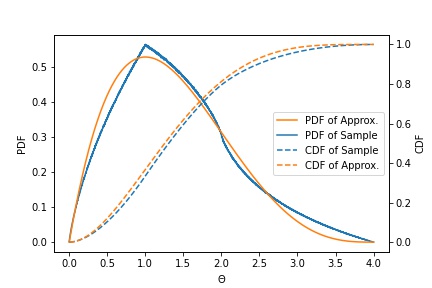}
    \caption{PDF \& CDF of orientation factor $\Theta$. Blue line shows the histogram of sampling from $(\cos{\theta},\phi/\pi,\psi/\pi,\cos{\iota})\sim U(-1,1)$ for $2.5\times 10^8$ times. The yellow line shows the polynomial approximation in \citet[Eq.3.11]{Finn1996PhysRevD.53.2878}. 
    Solid and dashed lines represent the PDF and CDF, respectively.}
    \label{fig:ThetaDF}
\end{figure}

According to the earlier paper\citep{Finn1993PhysRevD.47.2198}, these four orientation angles are combined into one orientation factor $\Theta$ by the following form:
\begin{equation}
    \label{eq:theta}
    \Theta = 2\left[ F_+^2\left(1+\cos^2{\iota}\right)^2 + 4F_\times^2\cos^2{\iota} \right]^{1/2},
\end{equation}
where the antenna patterns are:
\[ \begin{matrix}
F_+ = \frac{1}{2} \left( 1+\cos^2{\theta} \right) \cos{2\phi}\cos{2\psi} - \cos{\theta} \sin{2\phi}\sin{2\psi}, \\
F_\times = \frac{1}{2} \left( 1+\cos^2{\theta} \right) \sin{2\phi}\cos{2\psi} + \cos{\theta} \sin{2\phi}\cos{2\psi}.
\end{matrix} \]
In Fig.\ref{fig:ThetaDF}, we show the probability density function(PDF) $f(\Theta)$ and cumulative distribution function(CDF) $F(\Theta)$, which is directly sampled by MC method from four orientation angles $(\cos{\theta},\phi/\pi,\psi/\pi,\cos{\iota})\sim U[-1,1]$ of $2.5\times 10^8$ times, instead of using the polynomial approximation in \citet[Eq. 3.11]{Finn1996PhysRevD.53.2878}.

As for $D_L(z)$, we only assume the simulated sample of BBH systems are all at source redshift $z_s=10$ for simplicity.
The strong lensing effect is more predominant at high redshift \citep[see][Fig.3]{Piorkowska2013}, and the magnification PDF consisting of strong and weak lensing is not significantly changed after increasing redshift from 10 to 30
\citep[see][Fig.2]{Oguri2018MNRAS.480.3842O}.
Therefore, we could fix $z_s=10$ as a simple example\footnote{In fact, there have been some works discussing the detectability of BBH events with redshift up to $z\sim30$ \citep[][]{Nakamura2016PTEP.2016i3E01N, Koushiappas2017PhRvL.119v1104K}.} to illustrate the upper limit of the lensing effect on BHMF. 

Note that, we focus on the lensing effect which makes {\it relative} impacts on measured BHMF; the absolute number of the detected BBH event, which is determined by the BBH merger rate\citep[see][Eq.3]{Ding_2020}, is not related to this study. In our simulation, we constantly generate the random mock BBH systems until we collect 400,000 events for the coming study of adding  the lensing effect.

The last but most important parameters of our simulated BBH systems are $(m_1,m_2)$.
Here the mass of the heavier one $m_1$ is assumed to follow different BHMFs in each case respectively, then sampled by MC method.
We adopt three BHMF templates as examples: \textbf{Truncated} Power Law, Power Law with \textbf{Peak}, and \textbf{Broken} Power Law, as detailedly presented in Sec.\ref{sec:mass functions} and Fig.\ref{fig:IniBHMF} with parameters summarized in the "Initial" of  Tab.\ref{tab:parameters}, following \citet{LSC202105_GWTC2_pop}.
The second one $m_2$ are sampled simply from a uniform distribution as $m_2\sim U(m_\mathrm{low}, m_1)$ \citep{Ding_2020,LSC-GWTC1-Pop-2019ApJ...882L..24A}, since the BH mass ratio distribution is not concerned here. 
So later in our work, we will only select the heavier one $m_1$ for estimation to recover the parameters of BHMF, or reconstruct the BHMF, as we assume here. 
We note that the detailed initial shape of the BHMF is not very important for our purpose to explore the lensing effect, and our qualitative result will not be affected by a slight change of the PDF of $m_1$(same as \citet[]{Oguri2006PhRvD..73l3002O}). 
But, as we will present in Sec.\ref{sec:result}, the lensing effect on parameters does depend on how we describe the substructure of BHMF. 

Thus, we have got an unlensed catalog of BBHs. Their parameters including $(m_1,m_2)$ and $\mathcal{M}$, $(\theta,\phi,\psi,\iota)$ and $\Theta$, $z$ and $D_L(z)$ are all MC sampled from certain PDFs (BHMF, uniform, and fixing $z_s=10$).
As mentioned above, we haven't considered the lensing effect at this stage, i.e., this catalogue is just like in real analysis\citep[][]{LSC202105_GWTC2_pop}.
We then return to Eq.\ref{eq:SNRrho} to check whether we could detect these BBH-GW signals. 
Adopting the widely accepted detecting threshold $\rho_\text{th}=8$ in Eq.\ref{eq:SNRrho}, only about $1/3$ intrinsic events in our whole mocked universe with SNR $\rho>\rho_\text{th}$ could be detected.
So we need to handle this selection effect reasonably in Sec.\ref{sec:SelectionEffect} of estimation BHMF of our whole mocked universe by using only around $1/3$ BHs.

\subsection{Black Hole Mass Functions}
\label{sec:mass functions}

\begin{figure}
    \includegraphics[width=\columnwidth]{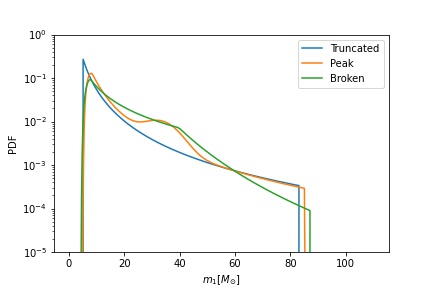}
    \caption{Probability Distribution Functions(PDFs) of three BHMF models, with parameters summarized in Table \ref{tab:parameters}. Blue, yellow, and green lines represent the Truncated Power Law, Power Law with Peak, and Broken Power Law, respectively.}
    \label{fig:IniBHMF} 
\end{figure}

Beyond the mass distribution of BHs observed in X-ray binaries \citep[see][Fig.1]{Kovetz2017PhRvD..95j3010K}, several latest coarse-grained models have been proposed in \citet{LSC202111_GWTC3_pop}, which are broadly consistent with their previously identified population\citep{LSC202105_GWTC2_pop}. 
So for simplicity, we follow the first three in \citet{LSC202105_GWTC2_pop},
along with the median value of their Bayesian parameter estimation summarized in Table.\ref{tab:parameters} for "Initial" and , as representative examples shown in Fig. \ref{fig:IniBHMF}.

\subsubsection{Truncated Power Law}

\textbf{Truncated} Power Law: 3 parameters, the \textit{simplest} model:
\begin{equation}
f_M\left(m_{1} | \alpha, m_\text{low}, m_\text{up}\right) =  
\begin{cases}m_{1}^{-\alpha}/C & ,m_\text{low}<m_{1}<m_\text{up} \\ 
0 & ,\text { otherwise }\end{cases}.
\end{equation}
This probability distribution function (PDF) of masses serves as the primary component for other mass models, which is simply modeled as a power-law function in an interval with two hard cut-offs. 
A twilight star with mass lower than the low-mass cut-off $m_\mathrm{low}$, is known to evolve into a neutron star or white draft. 
And the high-mass cut-off $m_\mathrm{up}$ denotes the theorized pair-instability gap (\citealt{Heger2002ApJ...567..532H, Woosley2015, Talbot2018ApJ...856..173T}; and \citealt{LSC2020ApJ...900L..13A...GW190521} Sec. 1.2 for a summary). 
Even though this simplest model has been confidently ruled out both in GWTC-2 \citep{LSC202105_GWTC2_pop} and GWTC-3 \citep{LSC202111_GWTC3_pop}, we still preserve it for our simulation, in order to show the primary effect on BHMF of lensing.

\subsubsection{Power Law with Peak}
Power Law with \textbf{Peak}: 7 parameters in $\vec{\theta}$, adding features beyond \textbf{Truncated} power-law, claimed to be \textit{the best-fit model}\citep{LSC202111_GWTC3_pop}:
\begin{align}
f_M\left(m_{1} | \vec{\theta} \right) &=  \left(1-\lambda_{\text {peak }}\right) \mathfrak{P}\left(m_{1} |-\alpha, m_\text{up}\right)+\lambda_{\text {peak }} G\left(m_{1} | \mu_{p}, \sigma_{p}\right), \notag \\ 
\vec{\theta} &= \left( \lambda_{\text {peak }}, \alpha, m_\text{low}, \delta_{m}, m_\text{up}, \mu_{p}, \sigma_{p} \right),
\end{align}
where $\mathfrak{P}$ (P in Fraktur) means a normalized power-law with a smooth function $S\left(m | m_\text{low}, \delta_{m}\right)$ at low mass end:
\begin{align}
    \mathfrak{P}\left(m |-\alpha, m_\text{up}\right) = \begin{cases}    \left. m^{-\alpha}S\left(m_{1} | m_\text{low}, \delta_{m}\right) \right/ C(m_\text{up}) & ,m<m_\text{up}\\
    0 & ,m\geq m_\text{up}\end{cases} 
\end{align}

\begin{equation}
\begin{aligned}
& C(m_\text{up}) = \int_0^{m_\text{up}} {m^{-\alpha}S\left(m_{1} | m_\text{low}, \delta_{m}\right) \cdot \text{d}m} \\
&S\left(m | m_\text{low}, \delta_{m}\right)= \begin{cases}0 & ,m<m_\text{low} \\
{\left[f\left(m-m_\text{low}, \delta_{m}\right)+1\right]^{-1}} & ,m_\text{low} \leq m<m_\text{low}+\delta_{m} \\
1 & ,m \geq m_\text{low}+\delta_{m} \end{cases} \\
&f\left(m^{\prime}, \delta_{m}\right)=\exp \left(\frac{\delta_{m}}{m^{\prime}}+\frac{\delta_{m}}{m^{\prime}-\delta_{m}}\right).
\end{aligned} \notag
\end{equation}
This model is based on ''Truncated'' with two modifications, consisting of a high mass Gaussian component $G\left(m_{1} | \mu_{p}, \sigma_{p}\right)$, and a smooth function $S\left(m_{1} | m_\text{low}, \delta_{m}\right)$ rising from 0 to 1 over the interval $(m_{\mathrm{low}},m_{\mathrm{low}}+\delta_{m})$, which makes the normalization constant $C$ more complicated.
The Gaussian component is motivated by a pile-up of BBH events before the pair-instability gap \citep[][]{Talbot2018ApJ...856..173T}.

\subsubsection{Broken Power Law}
\textbf{Broken} Power Law: 6 parameters in $\vec{\theta}$, providing an \textit{alternative} to the Power Law with \textbf{Peak} model:
\begin{equation}
\begin{array}{l}
f_M\left(m_{1} | \vec{\theta} \right) =  \begin{cases}m_{1}^{-\alpha_{1}} \cdot S /C & ,m_\text{low}<m_{1}<m_{\text {break }} \\
m_{1}^{-\alpha_{2}} \cdot S/C_2 & ,m_{\text {break }}<m_{1}<m_\text{up} \\
0 & ,\text { otherwise }\end{cases} \\
\qquad \vec{\theta} = \left( \alpha_{1}, \alpha_{2}, m_\text{low}, m_\text{up},b,\delta_m \right) \\
m_{\text {break }}=m_\text{low}+b\left(m_\text{up}-m_\text{low}\right), C_2 = C\cdot m_{\text {break }}^{\alpha_1-\alpha_2} .
\end{array}
\end{equation}

This model includes the same form of smooth function $S\left(m_{1} | m_\text{low}, \delta_{m}\right)$ in 'Power Law with Peak', and provides another power-law $m_{1}^{-\alpha_{2}}$ to describe the high mass component since certain mass $m_{\mathrm{break}}$, which 'could be though of as either a gradual tapering off, or a substructure of BHMF within the pair-instability gap'\citep{LSC202105_GWTC2_pop}.
And the normalization constant could be calculated based on the continuity of BHMF at $m_{\mathrm{break}}$.

\subsection{Lensing Effect}
\label{sec:lensing}

\begin{figure}
    \centering
    \includegraphics[width=\columnwidth]{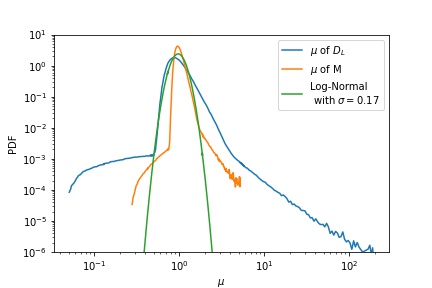}
    \caption{The distributions $P(\mu)$ of magnification $\mu$.
    The Luminosity Distance or Flux magnification $\mu_D$ (blue) are quoted from \citet[Figure 2, z=10]{Oguri2018MNRAS.480.3842O}.
    Mass magnification $\mu_m$ (yellow) are calculated via:$\quad z \rightarrow D_L(z)+ \mu_D \rightarrow D_L(\tilde{z}) \rightarrow \tilde{z} \rightarrow \mu_m$.
    Here we also present a Log-Normal Distribution (green) with $(m_0, \sigma)=(1, 0.17)$ in Eq.\ref{eq:Log-normal} as a comparison between lensing in Eq.\ref{eq:conv-dm} and uncertainty in Eq.\ref{eq:ConvUncertainty}.}
    \label{fig:mu}
\end{figure}

In this subsection, 
we take lensing effect into consideration.
We calculate the PDF of gravitational
lensing magnification according to the hybrid model described in~\citet[]{Oguri2018MNRAS.480.3842O}, as demonstrated in Fig.\ref{fig:mu}, the distribution of $P(\mu)$ at $z=10$, considering both strong and weak lensing.
Most of signals are magnified by a factor around 1 due to weak lensing, and about $10^{-3}$ of signals could be magnified tens to hundreds of times due to strong lensing.
\citet{Oguri2018MNRAS.480.3842O} assumed that multiple signals of GWs from strong lensing, are treated separately, i.e., identifying lensed signal pairs is difficult in reality, due to poor localization and lack of EM counterparts of binary BH mergers.
Thus, the normalization of $P(\mu)$ will  exceed unity, and we take this excess into account.

It is worth emphasizing that, the blue curve in Fig.\ref{fig:mu}, quoted from \citet[Fig. 2]{Oguri2018MNRAS.480.3842O} , is defined in optical (EM) observation.
This $\mu_{D}$ in $P(\mu_{D})$ actually describes the magnification of photon flux, not of the mass of BH we inferred from the GW signal. 
Following the geometric optical limit proposed in \citet[Eq.5]{Wang1996PhRvL..77.2875W}, it is the SNR $\rho$ (in Eq.\ref{eq:SNRrho}), which represents the square root of flux, that is directly amplified by a factor of $\sqrt{\mu_D}$:
\begin{equation}
\label{eq:shift rho}
    \Tilde{\rho}= \sqrt{\mu_D}\cdot \rho.
\end{equation}
Hereafter, we use tilde ($\tilde{\rho}$) to denote the lensed case.
Due to lensing magnification ($\mu_D>1$), 
part of the events which should have been intrinsically too faint to be detected, will now emerge above the detection threshold $\rho_\text{th}=8$. 
Conversely, de-magnification ($\mu_D<1$) will cause another part to submerge below $\rho_\text{th}$.

The magnification $\mu_m$ of mass $m_1,m_2$ estimated from BBH-GW signal, actually comes from the so-called '\textit{mass-redshift degeneracy}', as discussed in \citet{Dai2017PhRvD..95d4011D,Chen2019MNRAS.485L.141C}. 
Under Newtonian approximation \citep[see][ Eq.7 and Ref. there in ]{LSC2016AnP}, we could derive the chirp mass $\mathcal{M}$ from the observable, frequency $f$ and its derivative $\dot{f}:=\frac{\mathrm{d}f}{\mathrm{d}t}$ of GW signal, as an analytical form:
\begin{equation}
    \mathcal{M}=\frac{c^3}{G}\left( \frac{5}{96\pi^{(8/3)}}f^{-11/3} \frac{\mathrm{d}f}{\mathrm{d}t} \right)^{3/5}.
\end{equation}
But while the GW propagates along the null geodesic for a cosmological distance, the observed frequency (along with its derivative) will be redshifted by the expansion of the universe:$f_\mathrm{o}=f_s(1+z_s)^{-1}$,and $\Dot{f}_\mathrm{o}=\Dot{f_s}(1+z_s)^{-2}$.
So the ``Chirp Mass" we calculated is actually the observed ``Redshifted Chirp Mass" $\mathcal{M}_\mathrm{o}$, not the true source mass $\mathcal{M}_s$ \citep[see][Eq.5]{Chen2019MNRAS.485L.141C} :
\begin{equation}
    \mathcal{M}_\mathrm{o}=\frac{c^3}{G}\left( \frac{5}{96\pi^{(8/3)}}f_\mathrm{o}^{-11/3} \frac{\mathrm{d}f_\mathrm{o}}{\mathrm{d}t} \right)^{3/5}=\mathcal{M}_s(1+z_s).
\end{equation}
We could re-write it in terms of mass amplification $\mu_m$:
\begin{equation}
\label{eq:mass-redshift degeneracy}
    \Tilde{\mathcal{M}}=\frac{(1+z)}{(1+\Tilde{z})}\mathcal{M}:=\mu_m\mathcal{M}.
\end{equation}

Combined with this \textit{mass-redshift degeneracy}, we could deduce Eq.\ref{eq:shift rho} from Eq.\ref{eq:SNRrho}\&\ref{eq:mass-redshift degeneracy} that, enlarging the area or flux $\rho$ by $\sqrt{\mu_D}$ is simply equivalent to shifting the luminosity distance, as used in \citet{Dai2017PhRvD..95d4011D} and \citet[Sec 4]{LSC2021LensAPJ}:
\begin{equation}
\label{eq:shift DL}
    D_L(\Tilde{z})=\frac{D_L(z)}{\sqrt{\mu_D}},
\end{equation}
where we could inversely solve the distance-redshift ($D_L-\Tilde{z}$) relationship numerically to get the observed lensed redshift $\tilde{z}$ as follows:
\begin{align}
\label{eq:Dl-z}
    D_L(z)&=\left( 1+z \right) \frac{c}{H_0} \int_0^z \frac{\mathrm{d}z}{E(z)} ,\\
    \text{where } E(z)&=\sqrt{\Omega_\text{m} (1+z)^{-3}+\Omega_\Lambda} \quad
    \text{ under $\mathrm{\Lambda CDM}$ model. }
\end{align}

Thus we could calculate the true magnification distribution $P(\mu_m)$ of chirp mass $\mathcal{M}$ from the magnification distribution $P(\mu_D)$ of flux in \citet{Oguri2018MNRAS.480.3842O} via: $ \mu_D, z \xrightarrow[]{\text{Eq.}\ref{eq:Dl-z}} D_L(z)
\xrightarrow[]{\text{Eq.}\ref{eq:shift DL}} D_L(\tilde{z}) 
\xrightarrow[]{\ref{eq:Dl-z}} \tilde{z} \xrightarrow{\ref{eq:mass-redshift degeneracy}} \mu_m$. 
The PDF $P(\mu_m)$ is shown as yellow line in Fig.\ref{fig:mu}. 
Moreover, we plot a Log-Normal distribution function with $m_0=1,\sigma=0.17$ in Eq.\ref{eq:Log-normal} for comparison to noise, which will be introduced in the next section.

We obtain a lensed mock catalog by adding the lensing effect described by the PDF $P(\mu)$ of magnification $\mu$.
For each BBH-GW event in our mock catalogue in Sec.\ref{sec:catalog generation}, we could directly sample a magnification $\mu_m$ from $P(\mu_m)$, and multiply both $m_1,m_2$ and $\mathcal{M}$ by $\mu_m$, so as to imitate the observed mass magnified due to lensing.
The orbit of the BBH system $m_1,m_2$, compared to the lensing objects at cosmological distances, is so small that they can be seen as point sources to share the same magnification $\mu_m$.
Then we should re-check the SNR in Eq.\ref{eq:SNRrho} of each BBH-GW event due to magnification in Eq.\ref{eq:shift rho}, resulting in some events emerging above this threshold, and some sinking.

\subsection{Instrumental Noise}
\label{sec:noise}

In this section, we aim to simulate the uncertainties on the quantities (i.e. $D_L, \mathcal{M}, m_1$) inferred from the GW signals, approximated by a simple Log-Normal distribution following \citet[]{Ding_2020}.

As shown in Eq.\ref{eq:SNRrho}, the quantities measurable from GW signals are comprised of the SNR $\rho$, $D_L$, and $(1+z)\mathcal{M}$.
The official methodology of the LIGO-Virgo Collaboration is to use matched filter method and Bayesian Statistic to estimate the physical quantities(e.g.$m_1, D_L(z)$) of the BBH source \citep[see][for the first signal {\tt GW150914}]{LSC_GW150914Properties}.
The posterior distribution should have been normal and symmetric under large sample approximation \citep[see][Section 8.6.2]{rice2006Statistics}.
But in reality, parameters inferred from current instruments have asymmetric upper and lower uncertainty limits \citep[see][Figure 7 \& Table 4]{LSC202111GWTC3}). 

Following \citet[]{Ding_2020}, we assume that the quantities ($D_L, \mathcal{M}, m_1$) of our mock data obey the asymmetric Log-Normal distribution around the true values ($D_{L0}, \mathcal{M}_0, m_0$) with standard deviations $\sigma(D_L, \mathcal{M}, m_1) = (0.1950,0.1757,0.2484)$, by taking the results from \citet{Vitale2017PhRvD..95f4052V} at $z_s\in[6,20]$, which are consistent with \citet[Fig.7]{Shan2021MNRAS.508.1253S}  and \citet[Fig.12]{Maggiore2020JCAP...03..050M} for $D_L$.
Note that, almost all the papers, including \citet{LSC202111GWTC3, Vitale2017PhRvD..95f4052V}, show the 90\% confidence interval for parameters of quantile from 5th to 95th.
So we need to slightly transform the median values of uncertainty in \citet[LE of Fig 4,6,7]{Vitale2017PhRvD..95f4052V} to get our $\sigma(D_L, \mathcal{M}, m_1)$ (interval from $16\%$ to $84\%$) for Log-Normal distribution, in order to imitate the skewed distribution correctly.

Instead of the Normal distribution with error of $m_1-m_0=\epsilon\sim{N(0,\sigma)}$, the Log-Normal distribution $\ln{m_1}\sim{N(\ln{m_0},\sigma_m)}$ is essentially more like a conditional probability:
\begin{equation}
    \label{eq:Log-normal}
    P(m_1|m_0,\sigma_m)=\frac{1}{m_1\sqrt{2\pi}\sigma_m}\exp{\left[-\frac{\left( \ln{m_1}-\ln{m_0} \right)^2}{2\sigma_m^2}\right]},
\end{equation}
so as to $\ln{\mathcal{M}}\sim{N(\ln{\mathcal{M}_0},\sigma_\mathcal{M})},\ln{D_L}\sim{N(\ln{D_{L0}},\sigma_{DL})}$.
The parameters $\sigma$ in Log-Normal distribution are assumed to depend only on the instrument (as well as the data analysis pipeline), so we have them fixed, e.g. to 0.2484 for $m_1$ as above.

After adding the uncertainties, finally we get the mock data set in both unlensed and lensed scenarios and use Eq.\ref{eq:SNRrho} to test their detectability. 
We use those events beyond the detection threshold to estimate the BHMF.

\section{Parameter Estimation}
\label{sec:estimation}

In this section, we illustrate our methodology for BHMF estimation by using the unlensed and lensed data set above, in order to estimate the impact of lensing.

As said above, about 2/3 of events with $\rho<\rho_{th}$ could not be detected, 
but they must be taken into account in the BHMF estimation, known as Malmquist bias \citep[][]{Malmquist1922MeLuF.100....1M, Malmquist1925MeLuF.106....1M, Talbot2022ApJ...927...76T}.
We have to introduce a selection factor $\eta$ in Sec. \ref{sec:SelectionEffect} to 
estimate the undetected 2/3 by using the detected 1/3 of events.
Actually, it is the inverse process of the real instrumental threshold screening at the end of Sec.\ref{sec:catalog generation}.

In Sec. \ref{sec:likelihood}, we illustrate our fitting procedure following the algorithm in \citet{Ding_2020}.
We use Different Noise Realization to exhibit the distribution of the Maximum Likelihood Estimates of our pre-assumed BHMF models.

\subsection{Selection effect}
\label{sec:SelectionEffect}

To construct a model on selection effect, we follow \citet{Ding_2020} to introduce a selection factor $\eta$ and its reciprocal $y$ for a certain GW event. 
One could see in Eq.\ref{eq:SNRrho} that, when the inherent parameter $\mathcal{M}$ of the source is given, whether this signal can be detected, only depends on the orientation factor $\Theta$, as $D_L$ is fixed (to $z=10$).
So we define the probability for the detectable SNR $\rho$ to take on a value greater than threshold $\rho_{\text{th}}=8$ in $\Theta$-space at a certain given redshift $z$ and mass $\mathcal{M}$, as the selection factor $\eta$ with its reciprocal $y$:
\begin{align}
    \eta := P(\rho>\rho_{\text{th}}) =& P\left( \Theta> \Theta_{\mathrm{th}} \right) = 1-C_\Theta \left( \Theta_{\mathrm{th}} \right) \notag ,\\
    \label{eq:sf-factor_def}
    y:=\frac{1}{\eta},\quad \Theta_{\mathrm{th}} :=& \rho_{\text{th}} \left/{ \left[8\cdot \frac{r_0}{D_L(z)}\left( \frac{(1+z)\mathcal{M}} {1.2M_{\odot}}\right)^{\frac{5}{6}} \right] }\right. . 
\end{align}
where $C_\Theta(\Theta)$ denotes the CDF of the orientation factor $\Theta$ in Eq.\ref{eq:theta} as Fig.\ref{fig:ThetaDF}.
Therefore, the selection factor $\eta$ is a function of $\Theta_{\mathrm{th}}$, consisting of $\mathcal{M}$ and $D_L(z)$ (or $z$).
Again we use the interpolation function of CDF given by the sampling from 4 orientation angles in Eq.\ref{eq:theta}, instead of the polynomial approximation found in \citet{Finn1996PhysRevD.53.2878}.

Then we give an example to discuss the physical interpretation of $\eta$.
If we calculate $\eta:= P(\rho>\rho_{\text{th}})=0.2$ for a GW event, it means that we could detect only 20\% of GW signals with the same redshift and chirp mass $\mathcal{M}$ as this GW event in $\Theta$-space.
The SNR of other signals with lower orientation factors $\Theta$ and lower SNR $\rho$ will not be detected, but they need to be considered in BHMF estimation.
In order to correct this bias, we need to gain the event weight by $1/\eta:=y=5$ to take into account those missed GW events with the same mass but SNR $\rho$ below the threshold. Thus, the power index of $y:=1/\eta$ in the likelihood function Eq.\ref{eq:posterior} should be added.

Hereafter, note that we need the true $y_0$ calculated from the true value $\mathcal{M}_0, D_{L,0}(z_s)$ in Eq.\ref{eq:sf-factor_def}, instead of the detected or noised $\mathcal{M}, D_L(z_s)$ with uncertainties in our mock catalogue.
In fact, the latter represents the expected value $y_\text{E}$ of a certain distribution of $1/\eta$ due to the distribution of noised $\mathcal{M}, D_L(z_s)$.
This inequation $y_\text{E}\neq y_0$, is caused by the asymmetry of the Log-Normal random noise from both $(\mathcal{M}, D_L)$.
The distribution of $y:=1/\eta$ is, therefore, of course asymmetric (just like Log-Normal), causing the true value $y_0$, nor the mode value or the median value as in Normal distribution, not equal to the expected value $y_\text{E}$.

We want to find the difference between the true value $y_0$ we need and the expected value of $y_\text{E}$ we observed in our mock catalogue for unbiased estimation. 
Following \citet{Ding_2020} with a slight modification (of the denominator in Eq.\ref{eq:eta-equation} from 3 to 3.1) in our scenario, we assumed the distribution of $y$ follows another Log-Normal as its components $\mathcal{M},D_L(z_s)$ in Eq.\ref{eq:Log-normal}.
Then the expectation value of y can be calculated by its definition:
\begin{align}
    \text{assume:}\quad & \ln{y} \sim N\left( \ln{\left(y_0\right)}, \sigma_y^2 \right), \sigma_y :=  \ln{\left(y_\text{E}\right)} /3.1 \\ 
    \label{eq:eta-e}
    \Longrightarrow \quad &
    y_\text{E}:=\text{E}y = \int{y\cdot P(y|y_0) \mathrm{d}y} = y_0 \cdot \exp{\left(\frac{\sigma^2}{2}\right)}\\
    \label{eq:eta-equation}
    \Longrightarrow \quad  &
    \ln{\left(y_\text{E}\right)}=\ln{\left(y_0\right)} + \frac{1}{2\times3.1^2}\left( \ln{\left(y_\text{E}\right)} \right)^2.
\end{align}
Evidently, $y_\text{E} \neq y_0$.
Using the MC sampling, \citet[Fig.2]{Ding_2020} found that, this assumed Log-Normal could well describe the skewed distribution of $y:=1/\eta$, when given the true $y_0$. The $\sigma_y$ given by $y_\text{E}$ (as an assumption), is actually also a function of $y_0$, which could be solved from Eq.\ref{eq:eta-equation}.
Therefore, from the expected value $y_\text{E}$, which has been obtained by noised $\mathcal{M}$ and $D_L(z)$ in Eq.\ref{eq:sf-factor_def} in our catalogue, we could approximate the true value of selection factor $y_0$ following Eq.\ref{eq:eta-equation} as:
\begin{equation}
    y_0 = y_\text{E}\cdot\exp{\left[ -\left( \frac{\ln{y_\text{E}}}{3.1}\right)^2/2 \right]}.
    \label{eq:sf_factor0}
\end{equation}
We could see the Un-Lensed results (Blue line) in Fig.\ref{fig:corner1} of Sec.\ref{sec:result} that, the parameter estimation can well cover the true values of BHMF (Black square), confirming that not only our assumptions here---the selection effect--- but also our whole methods, are reasonable and available.

\subsection{Likelihood Function and Realization Approach}
\label{sec:likelihood}

In this section, we describe our fitting procedure.
As a predictive simulation, we aim at the distribution of best fits using Different Noise Realization.
In each realization, we use Maximum Likelihood Estimation (MLE) to get the best fit values.

Specifically, in each realization, we sample 1,000 events for each model of BHMF separately in Sec.\ref{sec:catalog generation}.
Then we sample from the $\mu$ distribution for the lensing effect, and sample from the Log-Normal distribution to imitate instrumental uncertainties on each GW event.
After considering the selection effect in Sec.\ref{sec:SelectionEffect}, we could estimate the parameters of BHMF 
in both Lensed/Un-Lensed scenarios, and compare them with initial set values to see how lensing biases.
Each realization produces a single point in parameter space, as one best-fit probability of BHMF.

Under the assumption of uniform prior, the posterior is proportional to the likelihood $\mathcal{L}$ as the product of the probabilities of all the events $f_M(m_0|\Vec{\theta})$ in Sec.\ref{sec:catalog generation}, each of which is convolved with the uncertainty $P(m_{1,i}|m_0)$ in Eq.\ref{eq:Log-normal} and selection factor $1/\eta_{0,i}$ in Eq.\ref{eq:sf_factor0} as the power index: 
\begin{align}
    \mathcal{L}(\Vec{\theta}|\Vec{m_1}) \propto 
    \prod_i^{\text{events}}{\left[\int_0^\infty {P(m_{1,i}|m_0)\mathrm{d}m_0\cdot f_M(m_0|\Vec{\theta})}\right]^{y_{0,i}}}
    \label{eq:posterior}.
\end{align}
Note that, we only considered the uncertainty caused by experimental noise in our estimation, which is embodied in the convolution (or the conditional probability) in $P(m_{1,i}|m_0)$.
This will be designated as Un-Lensed (estimation) model Eq.\ref{eq:ConvUncertainty} in Appendix.\ref{sec:convolution}.
And we should have taken into account another convolution term $P(\mu)$, implemented as the Lensed (estimation) model Eq.\ref{eq:conv3}.
The rationality will be detailed at the end of this section. 

The Maximum Likelihood Estimation(MLE), is used to get the 'best fit' parameters $\Vec{\theta}$ of BHMF when given a list of events $\vec{m_1}$ consist of its components $m_{1,i}$. 
This mode value $\Vec{\theta}$ is used as a single point to represent the whole posterior distribution of Bayesian parameter estimation as \citet[Figure 16\&17]{LSC202105_GWTC2_pop} did realistically. 

Then we do 1,000 realizations like this, in order to show the distribution of the best fits. 
For instance, in Fig.\ref{fig:corner1} and Tab.\ref{tab:parameters}, we can be 68\% certain (are 68\% sure) that the MLE value of the spectral index $\alpha$ in Different Noise Realization is limited to the range $2.34^{+0.11}_{-0.11}$ for lensed scenario, covering the true value of $a=2.40$.
Detailed results and discussions are presented in the next section.

Although in principle in Eq.\ref{eq:posterior}, we need to compare the estimation by using Lensed data under the Un-Lensed/Lensed model, rather than  Lensed/Un-Lensed data under the Un-Lensd model in our methodology. 
Since the data are all lensed, the consequence of ignoring the lensing effect can be calculated in Eq.\ref{eq:posterior}.
The difference lies in the acquisition of the reference truth value, that is Un-Lensed data to Un-Lensed model here, and Lensed data to Lensed model in reality.
But the introduction of one more convolution of lensing effect beyond the convolution of noise as Eq.\ref{eq:conv3} for a single event into Eq.\ref{eq:posterior}, would increase the computational effort to unmanageable. 
Therefore we simply use the Un-Lensed data under the Un-Lensed model to get the same un-biased estimation as Lensed data under Lensed model.
Or from another perspective, we use the Lensed/Un-Lensed data as a comparison, instead.

\section{Result and Discussion}
\label{sec:result}

In Sec.\ref{sec:data}, we generate our mock data for unlensed and lensed scenarios under 3 parameterized BHMF models.
Then in Sec.\ref{sec:estimation}, we illustrate the fitting procedure for recovering parameters of BHMFs using mock data under unlensed model.
Here in this Sec.\ref{sec:result}, results of parameters recovered are shown as distributions in Corner Plots Fig.\ref{fig:corner1},\ref{fig:corner2},\ref{fig:corner3}, along with their Quantile-Quantile(Q-Q) Plots Fig.\ref{fig:QQplot1},\ref{fig:QQplot2},\ref{fig:QQplot3} in Sec.\ref{sec:recovery}.
The numerical results are summarized in Tab.\ref{tab:parameters}, along with their p-values of the Kolmogorov-Smirnov(K-S) test.
Then in Sec.\ref{sec:reconstruction} Fig.\ref{fig:ConfidenceInterval_Noised0.84}, we reconstruct the BHMFs from these parameters distribution, to reveal the contribution of these biased parameters on BHMF.
Finally in Appendix.\ref{sec:convolution}, we turn to 'Convolution' for studying the lensing effect in another way.

\begin{table*}
    \centering
    \caption{Parameters summary of all three BHMF models. 
    'Initial' row in each sub-table shows the $(0.5,0.05,0.95)$ quantiles (median value and 90\% credible bounds) of  posterior distribution in GWTC-2\citep[see][Fig.16,17]{LSC202105_GWTC2_pop}.
    We quote the median value as our initial parameters for catalogue generation in Sec.\ref{sec:mass functions}.
    'Un-Lensed' and 'Lensed' rows show our estimation results, the same $(0.5,0.05,0.95)$ quantiles as in 'Initial' for comparing.
    The last row in each sub-table shows the p-value of the two-sample Kolmogorov-Smirnov test for Un-/Lensed scenarios.
    A low p-value($\leq0.05$ or 0.01 typically) indicates these two samples satisfy different distributions.
    The units of the parameters  characterizing the mass (i.e. $m_\text{up}, m_\text{low}, \delta_m, \mu_p, \sigma_p$) are all $M_{\odot}$, where $M_{\odot}$ is the Solar Mass, and no units for the other parameters.}
    \label{tab:parameters}
    \begin{tabular}{cclllllllll} 
    \hline
    Model & Parameters & $\alpha$ & $m_\text{up}$ & $m_\text{low}$ & $\delta_m$ & $\lambda_{\mathrm{peak}}$ & $\mu_p$ & $\sigma_p$ & $b_m$ & $\alpha_2$ \\
    \hline
    & Initial & 2.4 & 85 & 5 & & & & & & \\
    Truncated & Un-Lensed & $2.42_{-0.18}^{+0.18}$ & $84.14_{-9.66}^{+11.18}$ & $5.06_{-0.46}^{+0.40}$ &  & & & & &  \\
    & Lensed & $2.34_{-0.20}^{+0.17}$ & $90.08_{-11.68}^{+15.86}$ & $5.25_{-0.60}^{+0.50}$ & & & & & & \\
    K-S test & p-value & 0.0000 & 0.0000 & 0.0000 & & & & & &  \\
    \hline
    & Initial & $2.63_{-0.63}^{+0.79}$ & $86.22_{-12.98}^{+11.85}$ & $4.59_{-1.91}^{+1.39}$ & $4.82_{-4.04}^{+3.93}$ & $0.10_{-0.07}^{+0.14}$ & $33.07_{-5.63}^{+3.96}$ & $5.69_{-3.60}^{+3.78}$ & & \\
    Peak & Un-Lensed & $2.74_{-0.34}^{+0.40}$ & $89.02_{-15.56}^{+22.58}$ & $4.51_{-1.18}^{+1.55}$ & $5.32_{-3.50}^{+3.17}$ & $0.11_{-0.04}^{+0.05}$ & $32.82_{-3.41}^{+2.83}$ & $5.52_{-2.81}^{+2.91}$ &  & \\ 
    & Lensed & $2.64_{-0.33}^{+0.33}$ & $95.51_{-18.25}^{+25.72}$ & $4.59_{-1.29}^{+1.55}$ & $5.48_{-3.64}^{+3.05}$ & $0.11_{-0.04}^{+0.04}$ & $33.47_{-3.85}^{+3.45}$ & $5.72_{-2.94}^{+2.73}$ &  & \\
    K-S test & p-value & 0.0000 & 0.0000 & 0.2635 & 0.1339 & 0.0000 & 0.0000 & 0.0256 &  &\\
    \hline
    & Initial & $1.58_{-0.86}^{+0.82}$ & $87.14_{-12.22}^{+11.41}$ & $3.96_{-1.72}^{+1.94}$ & $4.83_{-4.30}^{+4.38}$ & & & & $0.43_{-0.12}^{+0.27}$ & $5.59_{-2.55}^{+4.08}$ \\
    Broken & Un-Lensed & $1.59_{-0.30}^{+0.26}$ & $81.72_{-20.22}^{+42.82}$ & $4.24_{-1.12}^{+1.55}$ & $4.87_{-3.30}^{+2.98}$ & & & & $0.41_{-0.13}^{+0.16}$ & $5.23_{-2.53}^{+2.51}$   \\ 
    & Lensed & $1.56_{-0.31}^{+0.29}$ & $91.12_{-27.24}^{+38.10}$ & $4.24_{-1.10}^{+1.56}$ & $5.16_{-3.60}^{+3.07}$ & & & & $0.37_{-0.10}^{+0.16}$ & $4.62_{-2.06}^{+2.13}$  \\
    K-S test & p-value & 0.0017 & 0.0000 & 0.8882 & 0.0028 & & & & 0.0000 & 0.0000 \\
    \hline
    \end{tabular}
\end{table*}

\subsection{Parameters Recovery}
\label{sec:recovery}

For each BHMF model, our method of 1,000 realizations give 1,000 samples to form a distribution (Fig.\ref{fig:corner1},\ref{fig:corner2},\ref{fig:corner3}) of the best fit(MLE) parameters $\vec{\theta}$ of BHMF. 
In addition to the quantiles of $(0.16,0.5,0.84)$ (representing the mean \& variance to some extent), we exhibit the K-S test (Tab.\ref{tab:parameters}) \& Q-Q plot (Fig.\ref{fig:QQplot1},\ref{fig:QQplot2},\ref{fig:QQplot3}) to compare these two Un-/Lensed samples, following Statistical Science.

The K-S test is typically used to determine whether two samples follow the same distribution.
A low p-value($\leq$0.05 or $\leq$0.01 typically) suggests we have $1-p = 95\% \text{ or } 99\%$ certainty that samples of this single BHMF parameter estimated from Lensed data follow a different distribution from Un-Lensed data, indicating the exact lensing effect on this certain BHMF parameter.

Then we display the Q-Q Plots (Fig.\ref{fig:QQplot1},\ref{fig:QQplot2},\ref{fig:QQplot3}) to show this lensing deviation explicity.
Q-Q Plot (Blue scatters) shows the equal cumulative probability quantiles as $y\sim x$ satisfying $F_1(y)=F_2(x):=\int_{-\infty}^x{f_2(x')\mathrm{d}x'}$, in order to show the different between two PDF $f_1(y)=f_2(x)\frac{\mathrm{d}x}{\mathrm{d}y}$.
For instance, if these two samples both follow Normal distribution $X_i\sim N(\mu_i,\sigma_i)$, and the Ordinary Least Squares(OLS) fit for the Q-Q scatters gives $y=kx+b$(Red line), then we conclude $\sigma_1=k\cdot\sigma_2,\mu_1=k\cdot\mu_2+b$.
In particular, the 45-degree line $y=x$(Black) marks where the two samples follow the same distribution $f_1(y)=f_2(x)$ precisely, as shown in the top left panels of Fig.\ref{fig:QQplot3}.

\subsubsection{Truncated Power Law}

\begin{figure}
    \includegraphics[width=\columnwidth]{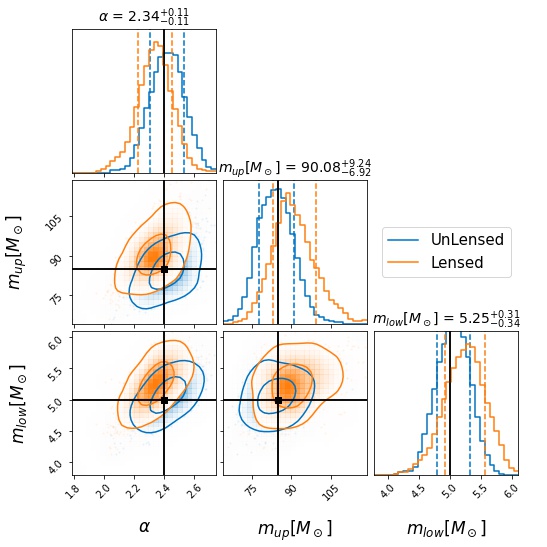}
    \caption{
    Corner Plot for Model 1 as Truncated Power-Law, consists of one- and two-dimensional distributions (histogram and contour) of the best-fit parameters in realization approximation.
    All the numerical results are summarized in the first sub-table in Tab.\ref{tab:parameters}.
    The black line marks the true value used for mock data generation.
    Blue and yellow lines represent Un-Lensed and Lensed scenarios, respectively.
    The contours represent $39.35\%, 86.47\%$ credible bounds as $1\sigma, 2\sigma$ in 2D contour.
    Two dashed lines represent the $(0.1587, 0.8413)$ quantiles as the boundary of $1\sigma$ in the 1D histogram.
    The values on the top of the histogram represent these two quantiles and the mean value for the Lensd scenario, instead of $(0.05, 0.95)$ quantiles in Tab.\ref{tab:parameters}.}
    \label{fig:corner1}
\end{figure}

\begin{figure*}
    \centering
    \includegraphics[width=\textwidth]{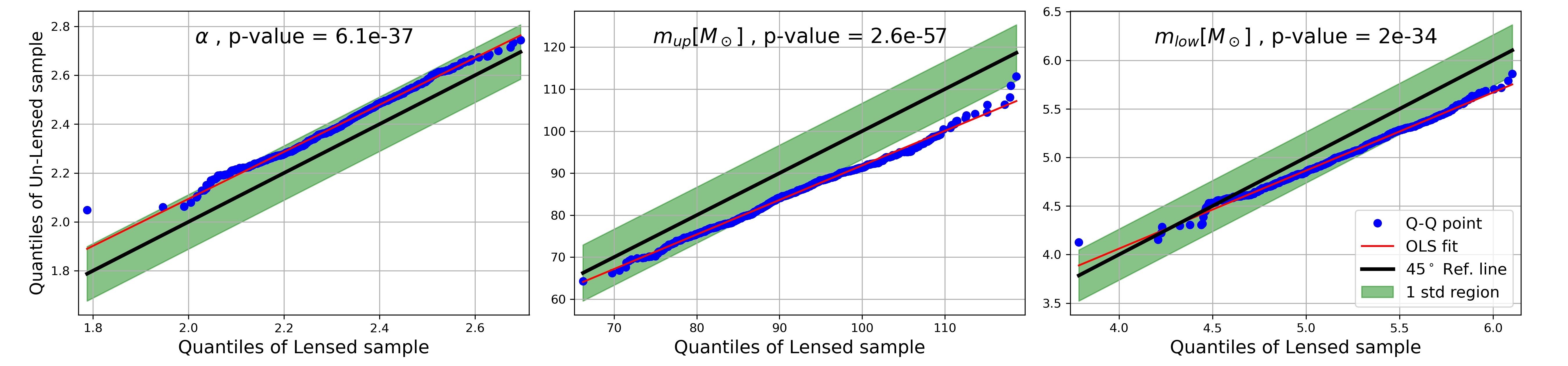}
    \caption{Quantile-Quantile(Q-Q) Plot(Blue scatters) of Un-/Lensed sample for some parameters in \textbf{Truncated} Power-Law.
    The Black lines mark the 45-degree reference line.
    The closer the Blue Q-Q scatters are to the $45^{\circ}$ Ref. line, the more similar the distribution of the two samples is, i.e., the smaller the effect of lensing on this parameter.
    The Red lines show the Ordinary Least Squares(OLS) fit for the Blue Q-Q scatters.
    The Green regions cover the interval of deviation from the $45^{\circ}$ Ref. lines for $1\sigma$(standard deviation) of Un-Lensed samples.}
    \label{fig:QQplot1}
\end{figure*}

For the simplest \textbf{Truncated} power-law spectral model, we could clearly see the lensing deviation from distributions in Fig.\ref{fig:corner1}.
The p-values of the K-S test in Tab.\ref{tab:parameters} are far below 0.0001, firmly suggesting lensing has an effect on the estimation of all these 3 BHMF parameters.
The Q-Q plots Fig.\ref{fig:QQplot1} indicate that all the effects are limited to around $1\sigma$(standard deviation, std).
That is, the distributions of the best-fit(MLE) BHMF parameters considering Lensing, are similar to the Un-Lensed distributions shifted by about 1 std, especially for the spectral index $\alpha$ in the first panel of Fig.\ref{fig:QQplot1}.

We conclude that for the \textbf{Truncated} model, the influence of lensing is limited to about $1\sigma$.
Similar results were presented in the traditional EM observation: \citet[Fig.12]{Charlotte2015ApJ...805...79M}, who study the effect of lensing on luminosity function at $z\sim8$.
Specifically, lensing slightly increases the upper and lower bounds $m_\mathrm{up}, m_\mathrm{low}$ of the power-law spectrum, which corresponds to the average magnification of the lens beyond 1 in Fig.\ref{fig:mu}, where $\langle \mu_{DL}\rangle=1.13,\langle \mu_{m}\rangle=1.021$.
And it slightly lowers the spectral index $\alpha$, which corresponds to the normalization, or overall dispersion and peak depression caused by the Full Width at Half Maximum(FWHM) of the magnification distribution $P(\mu_{m})$ over Dirac $\delta(1)$ function.

\subsubsection{Power Law with Peak}
\begin{figure*}
    \centering
    \includegraphics[width=0.95\textwidth]{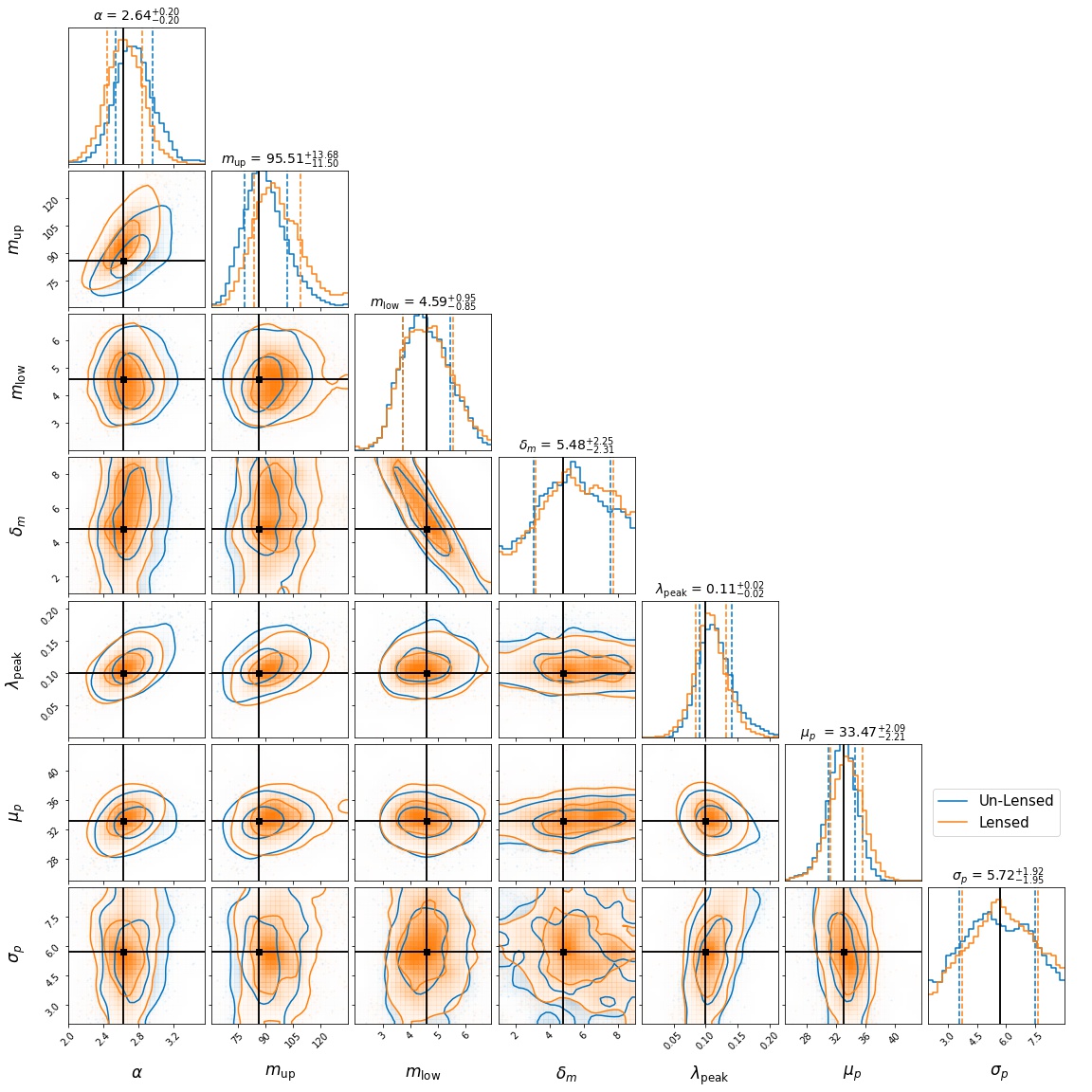}
    \caption{Same as Fig.\ref{fig:corner1}, but for Model 2 as Power-Law with Peak.}
    \label{fig:corner2}
\end{figure*}

\begin{figure*}
    \centering
    \includegraphics[width=\textwidth]{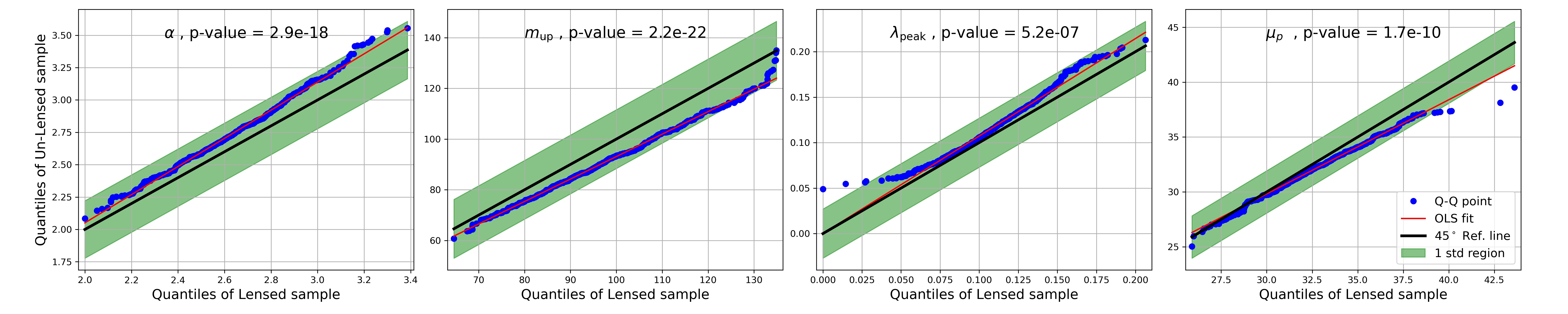}
    \caption{Same as Fig.\ref{fig:QQplot1}, but for Model 2 as Power-Law with Peak.
    Here we only show the parameters with a p-value$\leq0.0001$ in Tab.\ref{tab:parameters}, while others are so closed to the $45^{\circ}$ Ref. line like the left top panel in Fig.\ref{fig:QQplot3} that, they contain no more information than the p-value itself.}
    \label{fig:QQplot2}
\end{figure*}

As for the second power-law spectrum with Gaussian \textbf{Peak} model in Fig.\ref{fig:corner2},\ref{fig:QQplot2} and Tab.\ref{tab:parameters}, we find the more parameters, the weaker the effect of lensing on each single parameter. 
It becomes difficult to see the deviation directly from the distribution in Fig. \ref{fig:corner2}.
The p-values of the K-S test in Tab.\ref{tab:parameters} suggest that, for some parameters $(m_\text{low},\delta_m,\sigma_p)$, we really do not have sufficient confidence to reject the null hypothesis, i.e. we cannot perceive the lensing effect on the estimation of these parameters.
But the other p-values $(\alpha, m_\text{up}, \lambda_p, \mu_p)$ are low enough (as in Fig.\ref{fig:QQplot2}) to convince the lensing effect.

So we only exhibit Q-Q plots for these 4 parameters in Fig.\ref{fig:QQplot2}.
Almost all the Q-Q scatters fall inside the $1\sigma$ interval, implying that the lensing effect is not significant.
The deviation of spectral index and upper bound $(\alpha, m_\text{up})$ from about $0.2\sigma$ to $0.9\sigma$, is similar(but modest) to the previous simplest \textbf{Truncated} Power-Law model, and is also the result of the average lensing magnification and overall dispersion.
The deviation of $(\lambda_p, \mu_p)$ from about $0$ to $0.7\sigma$ is also similar to $(\alpha, m_\text{up})$ but far more slight.
Because they mark the substructure of BHMF at around $m\sim40M_\odot$, while lensing effect is more prominent at the high mass end of $m\sim90M_\odot$.

Outside the slight influence of lensing, the absence of deviations for $(m_\text{low},\delta_m,\sigma_p)$ can also be explained by this BHMF model itself.
This model is insensitive to these 3 parameters.
As the flat and peakless shape of the distributions estimated by real GW events \citep[see][Fig.16]{LSC202105_GWTC2_pop},
the posterior of them appear not so concentrated and they deviate from Gaussian significantly. 
So it is hard to find accurate mode values of Bayesian estimation as the MLEs here, leading to further deviations from Gaussian in the distribution of their MLEs under realizations, and less sensitive to lensing.
In fact, $(m_\text{low},\delta_m)$ in this BHMF model are introduced to show the non-empty low-mass gap\citep[see][Fig.7]{LSC202105_GWTC2_pop}.

\subsubsection{Broken Power Law}
\begin{figure*}
    \centering 
    \includegraphics[width=0.82\textwidth]{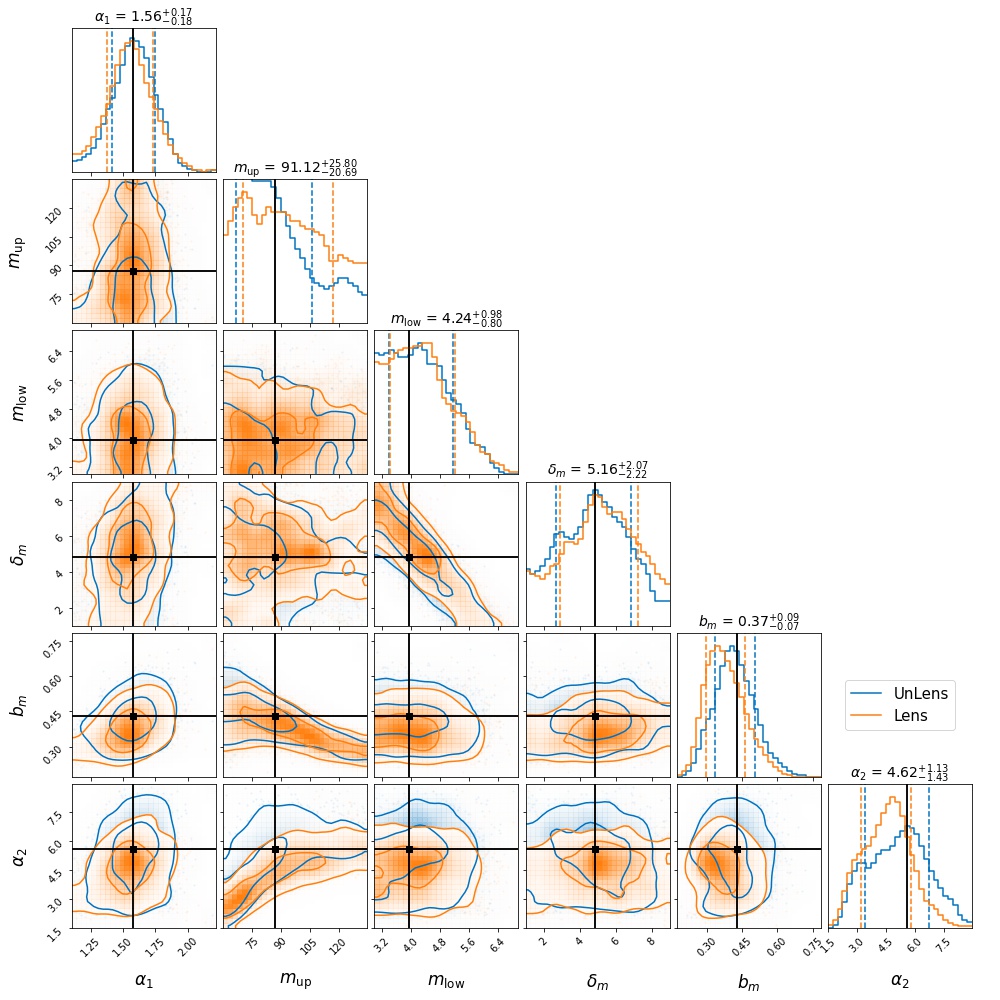}
    \caption{Same as Fig.\ref{fig:corner1}, but for Model 3 as Broken Power-Law.}
    \label{fig:corner3}
\end{figure*}

\begin{figure*}
    \centering
    \includegraphics[width=0.95\textwidth]{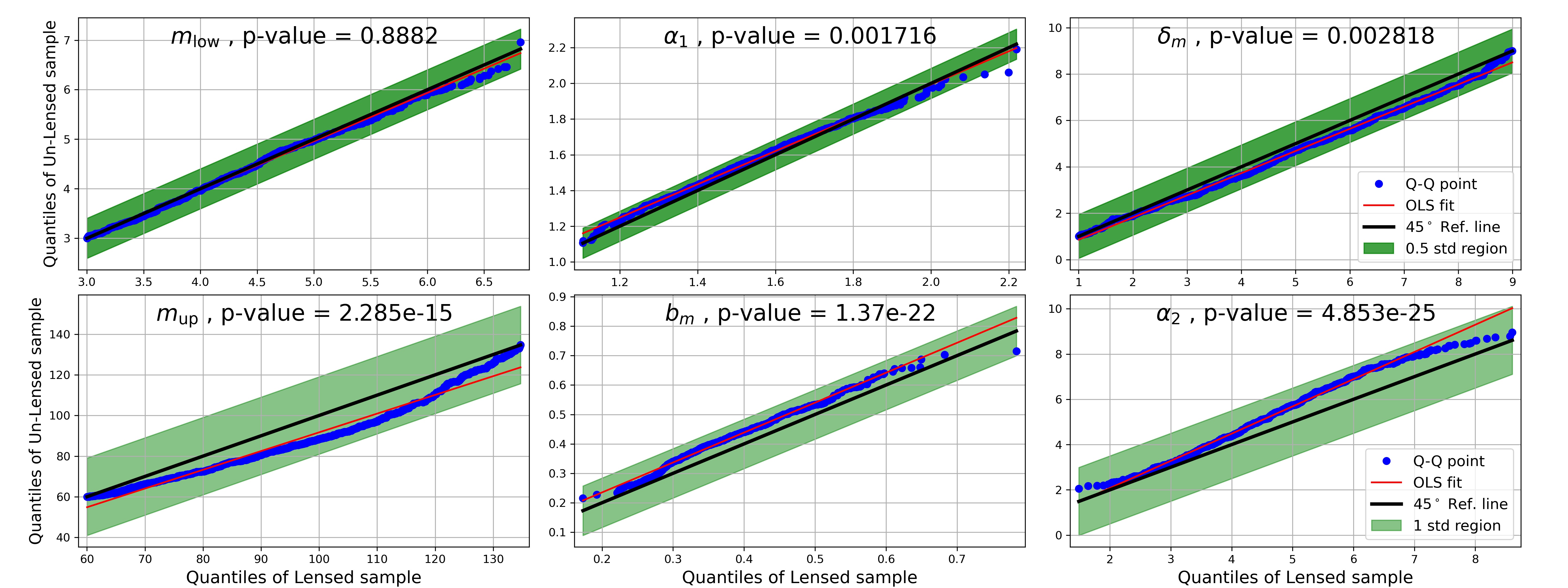}
    \caption{Same as Fig.\ref{fig:QQplot1}, but for Model 3 as Broken Power-Law.
    Here we exhibit all 6 parameters of this BHMF and have them rearranged, along with their p-values.
    The top 3 panels exhibit samples of parameters with non-zero p-values, so the Q-Q scatters are so close to the $45^{\circ}$ Ref. line that we cannot perceive the lensing effect.
    The bottom 3 panels exhibit the noticeable lensing effect on these parameters with nearly zero p-values.}
    \label{fig:QQplot3}
\end{figure*}

For the third double power-law spectral model Fig. \ref{fig:corner3}, \ref{fig:QQplot3} and Tab.\ref{tab:parameters}, the lensing effect is still very weak.
The distributions of some parameters $(m_\text{up}, m_\text{low}, \delta_m)$ deviate from the Gaussian clearly in 1-D histograms of Fig. \ref{fig:corner3}, similar to \citet[Fig.17]{LSC202105_GWTC2_pop}.
The non-Gaussianity indicates that this model does not depend strongly on these parameters, limiting the influence of lensing on them as for \textbf{Peak} model.

We show Q-Q plots for all parameters for this BHMF.
The large p-value of $m_\text{low}$ suggests no lensing effect on it, indicating that the Q-Q scatters are almost falling on the $45^{\circ}$ Ref. line in the top left panel of Fig.\ref{fig:QQplot3}.
Though the p-value $\sim 2\times10^{-3}$ of $(\alpha_1, \delta_m)$ are below the typical criterion 0.05 or 0.01, they are not low enough to display apparent deviation($\sim0.2\sigma$, specifically) from the $45^{\circ}$ Ref. line in the other 2 top panels.
Only the nearly zero p-values will indicate the noticeable lensing effect on these parameters in all the 3 bottom panels of Fig.\ref{fig:QQplot3}, about $0.5\sigma$, specifically.

These deviations of $(m_\text{up}, b_m, \alpha_2)$ are also reflected in the amplification on the high mass tail of the mass function and the dispersion on the low mass peak.
Furthermore, you could find that, these parameters, which will be affected by lensing, actually describe the substructure of BHMF at the high-mass end.

\subsection{BHMFs Reconstruction}
\label{sec:reconstruction}
\begin{figure}
    \centering
    \includegraphics[width=\columnwidth]{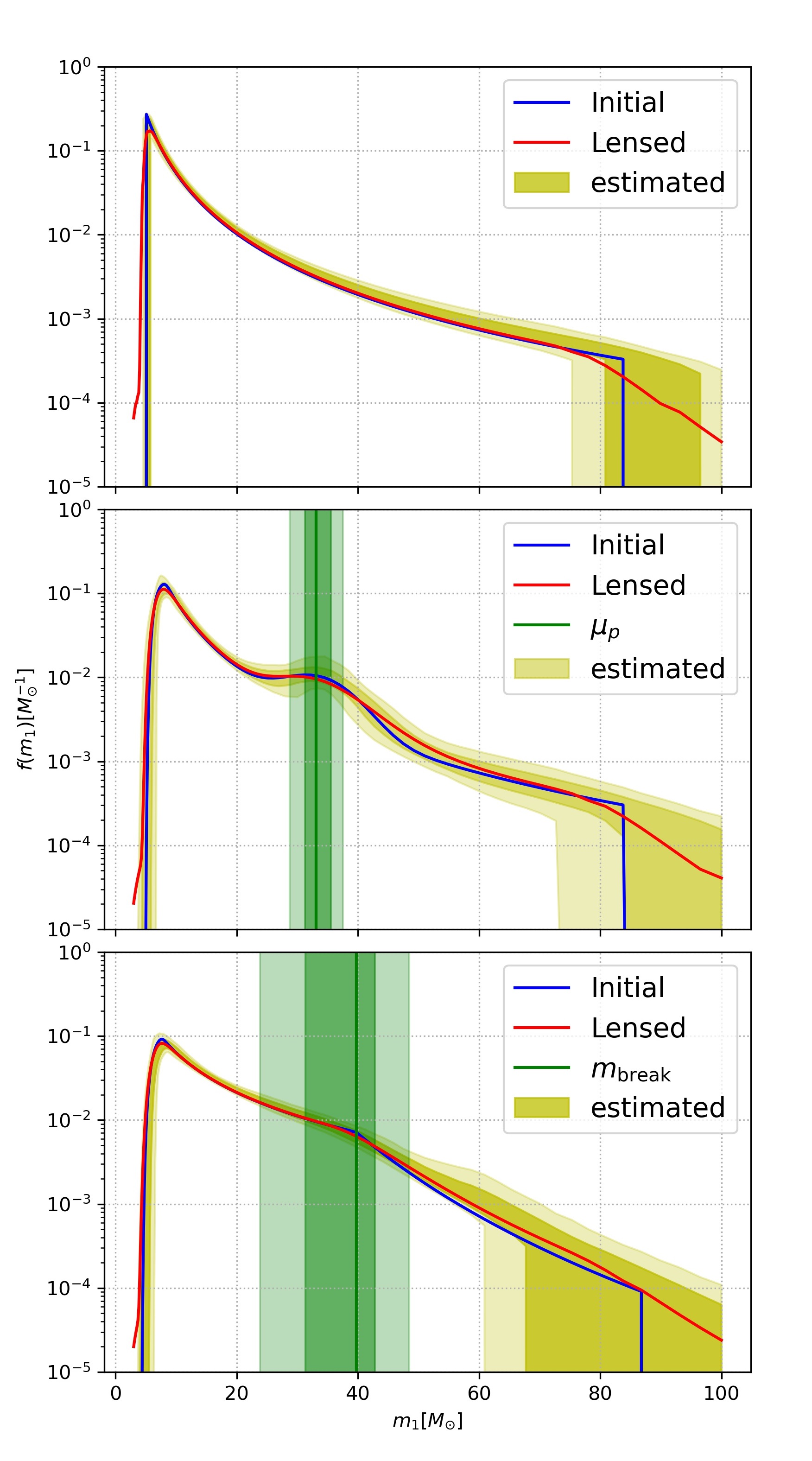}
    \caption{BHMF for the Truncated (top), Peak (middle), Broken (bottom) Power-Law.
    The blue lines are the Initial BHMF for generating our catalouge.
    The red lines are lensed BHMF convolved from Eq.\ref{eq:conv-m+miu}.
    The dark and light yellow region shows the 68\%, 95\% credible interval of BHMF reconstructed after estimation.
    The green shaded region shows the same interval of the feature at $m_1\sim 40M_\odot$ described by $\mu_p, m_\text{break}$ in Model 2,3 respectively, along with their initial values marked by the green line.}
    \label{fig:ConfidenceInterval_Noised0.84}
\end{figure}

We then try to reconstruct the whole BHMFs.
Each point of the MLE parameters from each realization in the corner plot above (e.g. Fig.\ref{fig:corner1}), leads to a single line of estimated BHMF (Fig.\ref{fig:ConfidenceInterval_Noised0.84}).
The dark and light yellow areas in Fig.\ref{fig:ConfidenceInterval_Noised0.84}, represent the $68\%,95\%$ confidence interval(i.e. $[0.1587, 0.8413]$ and $[0.0228, 0.9772]$ quantile) of BHMFs, respectively.
We also calculated Eq.\ref{eq:conv-m+miu} as the red line to show lensed BHMF without uncertainty as a comparison.

As seen from Figure \ref{fig:ConfidenceInterval_Noised0.84}, all  three estimated BHMF distributions using lensed data of 1000 realizations, can cover the preset unlensed distribution within 68\% confidence interval as $1\sigma$, indicating that lensing does not bias the estimation of the BHMF remarkably. 
The only deviation around $1\sigma$ is mainly reflected in the upper bound truncation for \textbf{Truncated} \& \textbf{Peak} models, and the power-law spectral index of the high mass part for model \textbf{Truncated} \& \textbf{Broken} models. 
It shows that the subtle influence of lensing is mainly reflected in the enlargement of the high mass tail, as well as the dispersion on the low mass peak due to normalization.

Moreover, here we consider why lensing affects different parameters on a different part of BHMF.
We deem that, for model 2, the index $\alpha$ has been well fitted by the more frequent low mass events, so lensing makes the upper end $m_\text{up}$ deviate more.
But for model 3, low mass events have been captured by the first index $\alpha_1$. 
So the prominent effect of lensing on the high mass tail will be reflected in the second index $\alpha_2$, more than in $m_\text{up}$.
Because the data used to estimate the parameter $\alpha_2$ at $m \sim 50M_\odot$, is higher in frequency than the number of events $m \sim 80M_\odot$, which determines the high mass end $m_\text{up}$.
As for the simplest model, the power-law model has only three parameters, so few so that the effect of lensing has to be clearly reflected in these simple parameters, including both the position parameter $m_\text{up}$ and the scale parameter $\alpha$ as the  main body in other two models.
But fortunately, this simplest model has been conclusively ruled out \citep[][]{LSC202105_GWTC2_pop, LSC202111_GWTC3_pop}, indicating that the impact of lensing will be shared by more parameters, especially those describing high-mass tail beyond main structure, indicating that the lensing effect is more imperceptible.

As a complement, we also tested BHs at lower redshifts, with lower uncertainty \citep[in][]{Vitale2017PhRvD..95f4052V} and fainter lensing effect \citep[in][]{Oguri2018MNRAS.480.3842O}.
For $z=3$, results are the same but fainter as for $z=10$ here. 
For $z=1$, the lensing effect is too faint to be distinguished, even compare to rather low instrumental uncertainty.
This means that as the redshift increases, the lensing effect grows much faster than the instrumental uncertainty.
So it is reasonable for us to only discuss the $z=10$ case, for the maximum influence of lensing as the upper limit.

\section{Summary}
\label{sec:summary}

We have investigated the impact of gravitational lensing on black hole mass function inference.
We use MC method to simulate the unlensed and lensed mock BBH catalogs, under 3 different intrinsic BHMF models for generation separately, i.e., truncated power, power law with peak, and broken power law.
We then estimate the parameters of BHMF from the mock data, taking instrumental uncertainty and selection effect into consideration.
We use the K-S test and Q-Q plot to discuss our parameter recovery, and convolution to discuss BHMF reconstruction.
Our results are summarized as follows.

\begin{enumerate}
 \item \textbf{Truncated}: the estimation of all 3 parameters $(\alpha, m_\text{up}, m_\text{low})$ could be biased by lensing. But the lensing effect is limited to around $1\sigma$(std.), indicating that the MLE will not be shifted drastically.
 The BHMF distributions reconstructed by these parameter distributions cover the true line in 68\% interval. And the true curve from $m\sim 40M_\odot$ goes almost close to the edge of 68\% interval(i.e. 0.1587 quantiles), as the deviation of both $(\alpha, m_\text{up})$ from lensing.
 \item \textbf{Peak}: some of the parameters $(m_\text{low},\delta_m,\sigma_p)$ are not affected by lensing. The others $(\alpha, m_\text{up}, \lambda_p, \mu_p)$ could be biased at around $0.5\sigma$.
 It is only at the high mass end $m\sim 80M_\odot$ of the BHMF distribution that, the true BHMF curve goes close to the edge of 68\% interval, indicating that $m_\text{up}$ is the most influential parameter for lensing.
 \item \textbf{Broken}: $(m_\text{low}, \alpha_1, \delta_m)$ are not affected by lensing. The others  $(m_\text{up}, b_m, \alpha_2)$ could be biased also at around $0.5\sigma$.
 The range where the true BHMF curve closest to the edge of the distribution interval is about $m\sim [40, 80]M_\odot$.
 This indicates that $\alpha_2$ is the most influential parameter in this model, while std. of $m_\text{up}$ is so large that it obscures the lensing deviation.
\end{enumerate}

We conclude that the lensing effect on the main structure of BHMF can be ignored, even for the GW events at rather high redshifts from the next-generation gravitational wave observatory.
The only considerable deviation is to elevate the high mass tail $m>50M_\odot$ of BHMF.
But this deviation of parameters of BHMF is still beneath the $1\sigma$ confidence interval, and strongly depends on the BHMF modeling.
It means lensing is mainly but only slightly reflected in the bias of a few parameters which capture the high-mass tail of BHMF.

\section*{Acknowledgements}

We thank Xi-Long Fan, and Yi Gong for useful discussions.
KL was supported by the National Natural Science Foundation of China (NSFC) under Grant Nos. 12222302, 11973034.
ZHZ was supported by NSFC Nos. 12021003, 11920101003, and 11633001 and the Strategic Priority Research Program of the Chinese Academy of Sciences, Grant No. XDB23000000.
LY was supported by JSPS KAKENHI Grant Number JP 21F21325.
\section*{Data Availability}

This theoretical study did not generate any new data.
 



\bibliographystyle{mnras}
\bibliography{refs} 




\appendix
\section{Lensing Convolution}
\label{sec:convolution}

\begin{figure}
    \centering
    \includegraphics[width=\columnwidth]{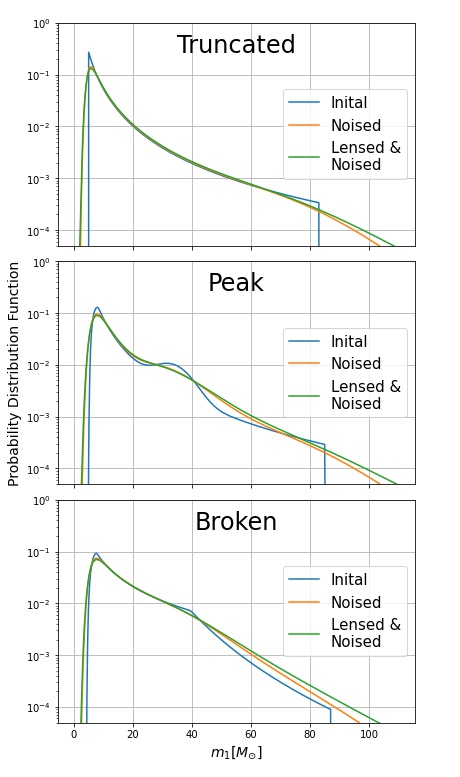}
    \caption{
    Initial (Blue) in Sec.\ref{sec:catalog generation}, unlensed \& noised (Yellow) Eq.\ref{eq:ConvUncertainty}, and Lensed \& noised (Green) Eq.\ref{eq:conv3} BHMF for each of three models.}
    \label{fig:ConvBHMF}
\end{figure}

In this appendix, we discuss the lensing effect in a more analytical way.

While the initial mass function $f_M\left( m_1 \right)$ and the magnification distribution $P(\mu_m)$ (replaced by $g(\mu)$ in this Appendix)are both given in Sec.\ref{sec:catalog generation} \& \ref{sec:lensing}, the distribution of lensed mass $\Tilde{m_1}=\mu_m\cdot m_1$ could be calculated as a distribution function of jointly distributed random variables, or a 'General Convolution'.
By using the methods in the textbook of classic statistics, we could derive the PDF of $f_{\Tilde{M}}(\Tilde{m})$:
\begin{align}
     \label{eq:conv-dm}
     f_{\Tilde{M}}\left( {\tilde m} \right) = \int_0^\infty {{\rm{d}}m \cdot f_M\left( m \right) g_{\mu}\left( {\frac{{\tilde m}}{m}} \right) \frac{1}{m}} \\
     \label{eq:conv-m+miu}
    = \int_0^\infty {{\rm{d}}\mu \cdot g_{\mu}\left( \mu \right) f_M\left( {\frac{{\tilde m}}{\mu }} \right) \frac{1}{\mu}},
\end{align}
where $(\mu,m)$ denotes $(\mu_m,m_1)$ for simplicity.
Remarkably, it is a little bit different from the commonly understood form of 'convolution', embodied in an extra factor of $\frac{1}{m}$ or $\frac{1}{\mu}$ in it by the first integration, which could be rechecked or understood as a requirement for normalization on $\tilde m$.
Hereafter we use Eq.\ref{eq:conv-m+miu}, since the PDF of $\mu_m$ is discretely calculated from \citet{Oguri2018MNRAS.480.3842O} and the initial PDF $f_M(m)$ is assumed to be continuous as the $f_M(m_1|\Vec{\theta})$ in Sec.\ref{sec:catalog generation}.

Then we 'convolve' the detection uncertainty of Log-Normal distribution $\ln{m_1}\sim{N(\ln{m_0},\sigma_m)}$ as conditional probability in Sec.\ref{sec:noise}, to get the observed BHMF:
\begin{align}
    \label{eq:ConvUncertainty}
    f_{\mathrm{unlens}}\left( {m} \right) & = \int_0^\infty {\mathrm{d}m_0\cdot P(m|m_0)f_M(m_0)} ,\\
    \label{eq:conv3}
    f_{\mathrm{lens}}({\tilde m}) & = \int_0^\infty {\mathrm{d}m_0\cdot P(\tilde m|m_0) \int_0^\infty {{\rm{d}}\mu \cdot g_{\mu}\left( \mu \right) f_M\left( {\frac{m_0}{\mu }} \right) \frac{1}{\mu}}}.
\end{align}
The overall effect that how gravitational lensing modifies the noised BHMF, are shown in Fig.\ref{fig:ConvBHMF}. 

As shown in Fig. \ref{fig:ConvBHMF}, under the non-negligible instrument uncertainty, the overall effect of lensing is rather weak. 
Both the peak at $\sim7M_\odot$ and the substructure at $\sim40M_\odot$ are depressed a lot due to convolution of lensing and uncertainty.
The dispersion at the high and low ends, indicate that we seem to detect more massive (mainly) black hole beyond the initial BHMF, in line with the amplification and diminution of gravitational lensing, as stated in \citet{Piorkowska2013, Ding2015JCAP...12..006D,yang2022MNRAS.509.3772Y, Oguri2018MNRAS.480.3842O}.
This high mass tail corresponds to the overestimation of truncated high mass end $m_\text{up}$.
For this reason, the normalization of 
$f_M(m)$ will inevitably compress peaks and substructures, resulting in underestimation of index $\alpha$ and making the details of BHMF difficult to be estimated.
Thus through convolution, we can explain the results of BHMF parameters estimation in Sec.\ref{sec:recovery},\ref{sec:reconstruction} to some extent.

But it seems the large instrumental noise $P(m_1|m_0)$ plays a more important role, instead of the lensing magnification $g\left( \mu \right)$.
And lensing appears to serve as another component of the uncertainty along with the $\sigma_m=0.2484$ instrumental noise.
More explicitly, it can be well approximated by the Log-Normal distribution of $\sigma=0.17$ shown in Fig.\ref{fig:mu} from our numerical test. 
Some works indeed used Log-Normal distribution to fit the main structure of the weak lensing, e.g. \citet{Dai2017PhRvD..95d4011D}, indicating that the lensing effect will dominate in the observed BHMF until the instrument uncertainty at $z=10$ is reduced far below lensing as $\sigma=0.17$.

The catalog of $m$ generated in Sec.\ref{sec:data}, is actually sampled from PDFs $f_M(m)$ for Un-Lensed and Lensed scenarios (Eq.\ref{eq:ConvUncertainty} and \ref{eq:conv3}) using three currently popular models (Fig.\ref{fig:IniBHMF}), respectively.
In the fitting procedure Sec.\ref{sec:estimation}, we only use Un-Lensed model with experimental noise as Eq.\ref{eq:ConvUncertainty} to reconstruct $f_M(m)$ in Eq.\ref{eq:posterior}.
The threshold screening $\rho_\text{th}=8$ and selection effect $y:=1/\eta$ are used to imitate the reality.
The final result in Sec.\ref{sec:result} for the lensed scenarios, in fact, comes from a fit to Eq.\ref{eq:ConvUncertainty} (Yellow lines in Fig.\ref{fig:ConvBHMF}) with the data sampled by Eq.\ref{eq:conv3} (Green lines in Fig.\ref{fig:ConvBHMF}).

The results of the convolution seem to provide a simple way to clarify the effect of lensing under the assumption of the large sample approximation.
However, in reality the limited samples, instrument error, and selection effect can easily make us overestimate the part of apparently massive BHs after lensing amplification. 
Hence, sampling under different noise realization approximations are still required, 
in recovering the parameters and reconstructing the BHMFs as in the main text, to study the lensing effect.


\bsp	
\label{lastpage}
\end{document}